# TUIMP: The Universe In My Pocket.
# Free astronomy booklets in all languages.

Grażyna STASIŃSKA[*1]

**Abstract.** TUIMP ([www.tuimp.org](www.tuimp.org)) is an international project to produce little astronomy booklets. These booklets, folded from just one sheet of paper, can be used in classrooms, at open public conferences, or during visits of observatories and planetariums. They are free to download from the internet, the only thing which is needed is a color printer (in absence of a printer, the booklets can also be directly consulted on line, even with just a mobile phone). The booklets are intended for children from nine years old and for anyone curious of astronomy. They are written in a simple language, amply illustrated, revised and translated by professional astronomers. So far, they are being published in six languages, others languages are to come. Everyone is invited to download the booklets and use them in their outreach activities.

### 1. Introduction

TUIMP stands for The Universe in My Pocket, in English. The characteristics of this outreach project are threefold. First, it does not require any funding and provides its products for free. Second, it allows anyone with an internet connection to download pdf files that can be printed and folded into small astronomy booklets. Third, it is open to all languages, in the hope of reaching populations that do not have much contact with astronomy.

### 2. Free

Even in the richest countries, outreach activities better reach their goals if they are free of charge. In less developed countries this is a necessity.

With TUIMP, all that is needed is a computer with an internet connection (and a printer).

### 3. Astronomy booklets

There are lots of sites for astronomy outreach on the internet. But people often like to have something to keep after a conference, a planetarium show or a school activity, so that they can remember or share their experience with their families.

TUIMP provides small 16-page booklets folded from one sheet of paper, free to download from the internet.

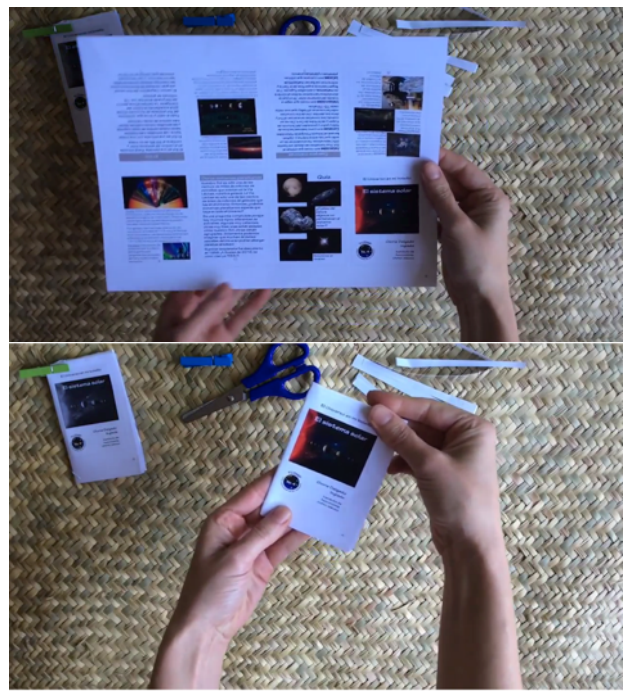

Fig. 1. How to fold a booklet.

### 4. In all languages

In countries with a large number of professional astronomers many persons are available to devote part of their time to astronomy outreach. In countries where astronomy is not much developed, the few professional astronomers are facing a large number of challenges and responsibilities, and developing original outreach activities is difficult for them.

Ready-made outreach internet sites are numerous in English but are likely scarce in the majority of languages, except a few in which an active astronomical community is working.

TUIMP provides material written by professional

[*1] LUTH, Observatoire de Paris, PSL, CNRS, UMPC, Université Paris Diderot, 5 place Jules Janssen, 92195 Meudon, France.
grazyna.stasinska@observatoiredeparis.psl.eu

astronomers that can be translated and used for free in any region of the World.

### 5. The target audience

The booklets are written with a wide audience in mind: children from nine years old but also any person curious about astronomy irrespective of their background. The language is simple, the texts are supported by numerous illustrations.

### 6. What makes the quality of TUIMP

The authors are professional astronomers. All the information and illustrations are taken from verified sources. There is an effort, in spite of very limited space, to put the astronomical material in a broader context and to touch on historical or sociological aspects. The scientific content is revised by external referees. To insure an accurate translation, the texts are translated by professional astronomers or astronomy students.

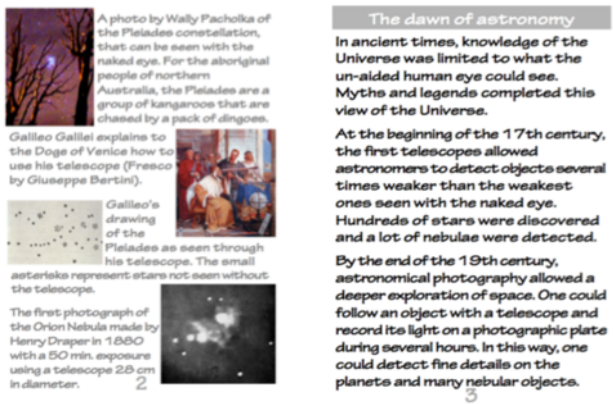

Fig 2. Two pages from the booklet "The invisible Universe".

### 7. The languages of TUIMP

Booklets are presently available in Albanian, English, French, Polish, Portuguese, and Spanish. Translations are being prepared into Armenian, Greek, Italian, Nahuatl, Persian and Russian.

Clearly many more languages are wished for, especially from African and Asian countries.

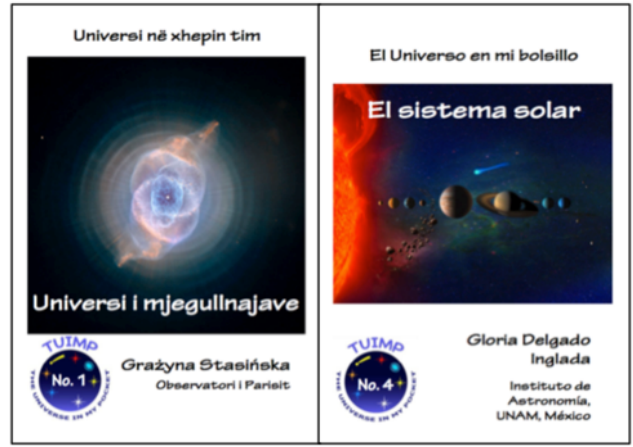

Fig. 3. The covers of two booklets, one in Albanese, the other in Spanish.

### 8. Where can the booklets be used?

They are perfect for activities in the classroom. They have much success when distributed after open public conferences on a related topic. They can be made available to visitors in planetariums and observatories. They also make nice little gifts for relatives and friends.

### 9. The TUIMP team

Initially, the team was composed of just a few persons from different parts of the world: Fabricio Chiquio Boppré (Brazil), Gloria Delgado Inglada (Mexico), Mimoza Hafizi (Albania), Dorota Kozieł-Wierzbowska (Poland), Stan Kurtz (USA), Grażyna Stasińska (France), Natalia Vale Asari (Brazil). The team is growing, with astronomers from Greece, Italy, Iran, Armenia joining in. All the participants – including the webmaster – give their time for free.

New authors and translators are welcome to join us. For this they can visit www.tuimp.org and write to us using the contact form to be found on our site at http://www.tuimp.org/pages/about.

### 10. Conclusions

The TUIMP project has started off well. It needs now to find its public. It is crucial to introduce it to potentially interested people all over the world, i.e. educators, children and teen agers, students, scientific journalists, astronomy aficionados and anyone interested in science. For this, the international astronomical community can be of great help.